\documentclass[conference]{IEEEtran}
\IEEEoverridecommandlockouts
\usepackage{cite}
\usepackage{amsmath,amssymb,amsfonts}
\usepackage{algorithmic}
\usepackage{graphicx}
\usepackage{textcomp}
\usepackage{xcolor}
\usepackage{listings}
\usepackage[breaklinks=true]{hyperref}
\usepackage{comment}
\usepackage[switch]{lineno}
 \usepackage{inconsolata}

\def\BibTeX{{\rm B\kern-.05em{\sc i\kern-.025em b}\kern-.08em
    T\kern-.1667em\lower.7ex\hbox{E}\kern-.125emX}}

\newcommand{\dd}[2]{\frac{\partial #1}{\partial #2}}
\newcommand{\paren}[1]{\left( #1 \right)}
\begin{document}
%\linenumbers 
%\IEEEpubid{\makebox[\columnwidth]{ This work has been submitted to the IEEE for possible publication. 
%Copyright may be transferred without notice, after which this version may no longer be accessible.\hfill} \hspace{\columnsep}\makebox[\columnwidth]{ }}
\IEEEpubid{\begin{minipage}{\textwidth}\ \\[12pt]
  This work has been submitted to the IEEE for possible publication. \\
  Copyright may be transferred without notice, after which this version \\
  may no longer be accessible.
\end{minipage}} 

\title{Optimizing the Weather Research and Forecasting Model
with OpenMP Offload and Codee\\
}
\author{
\IEEEauthorblockN{Chayanon (Namo) Wichitrnithed
\IEEEauthorrefmark{1}, 
Woo-Sun-Yang
\IEEEauthorrefmark{2}, 
Yun (Helen) He
\IEEEauthorrefmark{2}, 
Brad Richardson
\IEEEauthorrefmark{2}, 
Koichi Sakaguchi
\IEEEauthorrefmark{3}, \\
Manuel Arenaz
\IEEEauthorrefmark{4},
William I. Gustafson Jr.
\IEEEauthorrefmark{3},
Jacob Shpund
\IEEEauthorrefmark{5},
Ulises Costi Blanco
\IEEEauthorrefmark{4},
Alvaro Goldar Dieste
\IEEEauthorrefmark{4}
}
\IEEEauthorblockA{
\IEEEauthorrefmark{1}
Oden Institute for Computational Engineering \& Sciences\\
The University of Texas at Austin, Austin, TX 78712\\
Email: namo@utexas.edu}
\IEEEauthorblockA{\IEEEauthorrefmark{2}
Lawrence Berkeley National Laboratory\\
}
\IEEEauthorblockA{\IEEEauthorrefmark{3}
Pacific Northwest National Laboratory\\
}
\IEEEauthorblockA{\IEEEauthorrefmark{4}
Appentra Solutions S.L.\\
}
\IEEEauthorblockA{\IEEEauthorrefmark{5}
The Hebrew University of Jerusalem\\
}
}

\maketitle

\begin{abstract}

Currently, the Weather Research and Forecasting
model (WRF) utilizes shared memory (OpenMP)
and distributed memory (MPI) parallelisms. To take
advantage of GPU resources on the Perlmutter supercomputer at NERSC, we port
parts of the computationally expensive routines of the Fast
Spectral Bin Microphysics (FSBM) microphysical scheme to NVIDIA GPUs using
OpenMP device offloading directives. To facilitate this process, we explore a workflow for optimization which uses both runtime profilers and a static code inspection tool Codee to refactor the subroutine.  We observe a 2.08x overall speedup for the CONUS-12km 
thunderstorm test case.
\end{abstract}


\section{Introduction}

The Weather Research and Forecasting (WRF) model is an atmospheric model written in Fortran that solves the 3D Euler equations using finite differences \cite{Skamarock2021}. It is able to predict state variables such as temperature, humidity, and winds. Clouds within weather models like WRF are parameterized using a combination of cumulus and microphysics parameterizations. 

\begin{figure}[b]
    \centering
    \includegraphics[width=0.85\linewidth]{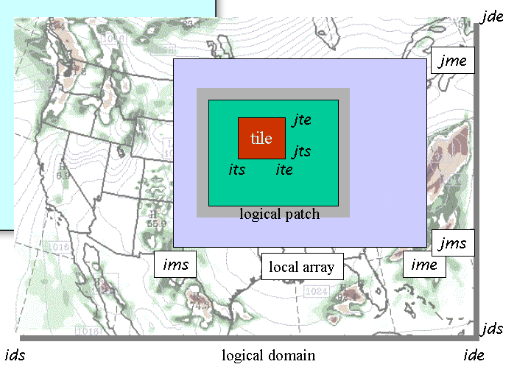}
    \caption{WRF decomposition layer. Diagram from \cite{MichalakesUnknown-db}.}
    \label{fig:tiles}
\end{figure}

WRF currently supports parallel computation only through
domain decomposition (MPI) and shared
memory (OpenMP) within each domain in the horizontal dimensions. This is illustrated in Figure \ref{fig:tiles}. The overall grid with array ranges (\verb|jds:jde,ids:ide|) is partitioned into rectangular  \emph{patches} with ranges (\verb|jms:jme,ims:ime|), each assigned to an MPI task. Within each patch, work can be further split into \emph{tiles} with ranges \verb|(jts:jte,its:ite)| and distributed among OpenMP threads.

One particularly expensive microphysics parameterization is the Fast Spectral-Bin Microphysics (FSBM) scheme, which calculates grid-resolved cloud condensate variables \cite{Khain2004-vs,shpundSimulatingMesoscale2019}. The formulation of FSBM uses discrete size intervals (bins) for cloud droplets and raindrops. This discretization can be extended from 33 to a few hundreds bins in order to improve convergence toward a more precise solution. 
The computational cost of this technique scales quadratically with the number of bins per grid point, making it an attractive portion of the code to port to GPUs. 
Doing so would also provide guidelines for a future rewrite of the scheme that is fully optimized for the GPU.

%Manuel
%Manuel[for Manuel: I added a paragraph from the Codee Training announcement about Codee in Section 5A. I think it can be split into a brief intro here with cite to Codee and more details in 5A. Or probably better to just have the Codee intro in Section 5A entirely, and in Section 1 maybe say something about the lack of such tools?]
%Manuel
%Manuel [Codee + profiling workflow for optimizing legacy code]
%Manuel
The development of large HPC software packages is a complex, time-consuming process even for experienced programmers, and WRF is not an exception. Codee\cite{Codee} is a new static code analysis tool that enables a more systematic, predictable approach to the modernization and optimization of Fortran/C/C++ codes. Designed as a complement to profilers and compilers, it facilitates modernization of legacy code, porting to GPUs using OpenMP/OpenACC directives, and automated testing in CI/CD frameworks. 
% comment {feels some repetion as in Section 5A. Commented out by Helen: Important features include its ability to identify opportunities in Fortran/C/C++ code leveraging the Open Catalog of Best Practices for Modernization and Optimization\cite{OpenCatalog}, insert OpenMP and OpenACC directives to exploit CPU and GPU parallelism, as well as automatically rewrite code to enforce Fortran modernization best practices.}

%Manuel
%Manuel[add a paragraph here to say what we did in this paper, somewhat rewording the abstract, and cite references to NERSC, Perlmutter, and Codee]: 
%Manuel
In this paper, we ported parts of the computationally expensive routine FSBM to NVIDIA GPUs on the National Energy Research Scientific Computing Center (NERSC)\cite{NERSC} supercomputer Perlmutter\cite{Perlmutter} using
OpenMP device offloading directives. To facilitate this process, we explored a workflow for optimization. Runtime profilers and a static code inspection tool, Codee\cite{Codee}, are used to refactor the subroutine. 

The organization of this paper is as follows: 
Section \ref{sec:related} lists  some of the GPU parallelization efforts that have been performed on WRF and related models. 
Section~\ref{sec:fsmb-routine} describes the FSBM scheme and its computational structure. 
Section \ref{sec:setup} describes the system configuration and compilers used. Section \ref{sec:approach} describes the overall approach of using OpenMP and Codee. Section \ref{sec:implementation} goes into the details of the profiling and optimization process. Section \ref{sec:multiple} presents additional performance evaluations and verification of the code. Finally, Section \ref{sec:conclusion} discusses the performance results and future work.

\section{Related work}\label{sec:related}
\begin{comment}
[this section is a little weak. citing more recent work is good too. 
Found: 
-- 2021: https://www.cisl.ucar.edu/sites/default/files/2021-10/Abdi-Acceleration%20of%20WRF%20on%20the%20GPU.pdf
-- 2023: https://www.hpcwire.com/2023/04/18/meet-tempoquest-a-startup-bringing-weather-code-into-the-accelerated-era/ 
-- 2015: https://gmd.copernicus.org/preprints/7/8031/2014/gmd-2014-180-manuscript-version3.pdf
-- 2022: https://www.researchgate.net/publication/367554316_Enhancement_of_WRF_Model_Using_CUDA
-- Stan Posey is the Nvidia climate GPU person, Found a talk in 2021: https://www.cisl.ucar.edu/sites/default/files/2021-10/Posey-NVIDIA_MC8_AS_Update.pdf
-- 2020, Stan Posey: https://www.researchgate.net/publication/356786174_Performance_Evaluation_of_the_Weather_Research_and_Forecasting_WRF_Model_on_the_DOE_Summit_Supercomputer
]
\end{comment}
Previous efforts to incorporate GPU acceleration in weather physics routines include \cite{Michalakes2008-cd}, \cite{Iacono2014-lb}, and \cite{Kim2021-bk}. The recent work \cite{Kim2021-bk} offloads the single-moment 6-class microphysics scheme (WSM6) within the Model for Prediction Across Scales (MPAS) using OpenACC. They performed several  optimizations such as subroutine inlining and packing, and reported a speedup of 2.38x on a single Tesla V100 GPU over 48 MPI tasks (3.00 GHz Intel Xeon Gold 6136). The dataset used consisted of 163,842 cells in the horizontal dimensions.

In \cite{Iacono2014-lb}, radiation routines in WRF are ported to GPUs through CUDA Fortran, where multiple blocks of profiles are distributed among threads.  Multiple optimizations were performed such as array padding and restructuring for coalesced accesses, demonstrating a speedup of 12.18x on an NVIDIA Tesla M2070-Q GPU over a 2.6 GHz Intel Xeon E5-2670 processor on a 4380 \texttimes\ 29 grid. 

In \cite{Michalakes2008-cd}, a GPU version of the single-moment 5-tracer microphysics scheme (WSM5) in WRF was created by converting the original Fortran code to CUDA C. The grid points are distributed in a coalesced, one-thread-per-vertical-column fashion. They reported a WSM5 speedup of 20x on a NVIDIA 8800 GTX GPU compared to a 2.80 GHz Pentium-D CPU processor on a grid with 115,000 cells. Another CUDA C implementation can be seen in \cite{Huang2015-mb}, where the Yonsei University planetary boundary layer (YSU-PBL) scheme is accelerated. After several CUDA optimizations and loop restructuring, the scheme observed a speedup of 193x on an NVIDIA Tesla K40 GPU compared to an Intel Xeon E5-2603 processor. 

There have also been efforts to port WRF to run entirely on GPUs. In \cite{PoseyUnknown-lf}, the NVIDIA-led WRFg is tested on the CONUS-2.5km test case with limited physics options. They report a speedup of 4x on an NVIDIA V100 GPU versus a 2.2 GHz Intel Xeon E5-2698 processor. 

Finally, in \cite{Abdi_undated-ek}, a proprietary port of WRF called AceCAST-WRF by TempoQuest is done through OpenACC and CUDA. They observed a speedup of 5x to 7x for the CONUS-2.5km case running on 4 NVIDIA P100/V100 GPUs versus 32 CPUs. Additional benchmarks are described in \cite{Sever2020-ld}.

\section{The FSBM routine}\label{sec:fsmb-routine}
Parameterization of microphysical processes can be largely divided into two categories, \emph{bin schemes} and \emph{bulk schemes}. In the more commonly used bulk schemes, the particle size spectrum is assumed to be represented by an analytic function (typically a gamma or Gaussian distribution), and this function is evolved over time by calculating the first few moments.

In contrast, bin schemes like FSBM divide particle size distributions into discrete size or mass (bins), and computation is done for solving equations explicitly for every bin  (see Figure \ref{fig:bulk}). This results in more equations to be solved at each grid point than bulk schemes. The FSBM scheme in particular uses 33 bins, and the predictive equation of the $k^{th}$ bin of particle type $i$ is as follows:

\begin{figure}
    \centering
    \includegraphics[width=\linewidth]{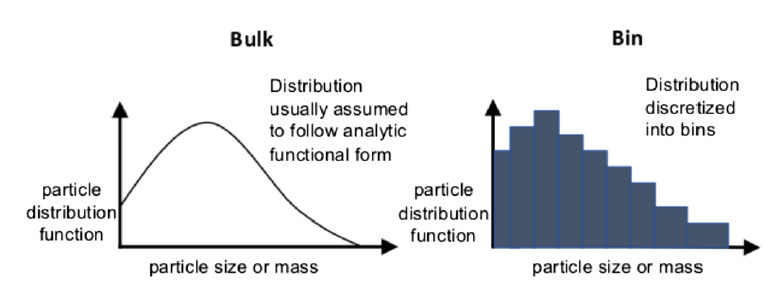}
    \caption{Comparison of bulk and bin microphysics schemes. Image from \cite{Morrison2020-zn}.}
    \label{fig:bulk}
\end{figure}
\begin{align}
    \dd{f_{ik}}{t} + \dd{\,uf_{ik}}{x} + \dd{\,(w-V_{tik})f_{ik}}{z} 
    = \sum_p \paren{\dd{f_{ik}}{t}}_{p},
\end{align}
where $V_{tik}$ is the terminal velocity, $u$ and $w$ the horizontal and vertical velocity components, and the right hand side the sum of the changes due to nucleation, condensation/evaporation, deposition/sublimation, collisions, freezing/melting, and breakup \cite{Khain2004-vs}.

This FSBM scheme is implemented in WRF as the subroutine \verb|fast_sbm| which contains further calls to other processes such as condensation, sedimentation, and collisions. To confirm that FSBM is indeed a hotspot, we first used GNU gprof to quickly gain a rough estimate of the top few hot spots, aggregating the output from all MPI cores. Due to load imbalances for FSBM, we further select a particular MPI task and annotate these subroutines with NVTX markers. We then used the NVIDIA Nsight Systems profiler\cite{nsight_systems} to compute the time contribution. This was done using 16 MPI tasks running on the CONUS-12km test case which contains 425 \texttimes\ 300 \texttimes\ 50 grid points, and the results are shown in Table \ref{table:gprof}.
\begin{table}[h!]
\begin{center}
\begin{tabular}{ c c c  }
\hline
 Routine & gprof & Nsight Systems  \\ 
 \hline
 \verb|fast_sbm| & 51.39 & 77.07 \\
 \verb|rk_scalar_tend| & 28.07 & 10.15 \\
 \verb|rk_update_scalar| & 6.361 & 1.504 \\
 \hline
\end{tabular}
\end{center}
\caption{Time contribution (\%) of the top few hotspots in WRF as reported by gprof. The corresponding Nsight Systems measurements are done for a single task.}.
\label{table:gprof}

\end{table}

While there are discrepancies between gprof and Nsight Systems, these measurements are sufficient for the purpose of choosing optimization targets.

Further measurements of the subroutines inside \verb|fast_sbm()| with gprof show that the collision-coalescence routine (\verb|coal_bott_new|) occupies a large percentage of computational time, making it a good candidate for optimization. 
\begin{comment}
\begin{table}[h!]
\begin{center}
\begin{tabular}{ c c c  }
\hline
 Process & Percentage of FSBM & Overall percentage  \\ 
 \hline
 Collision & 67.15 & 43.51 \\
 Freeze &  & \\
 Condensation &  & \\
 \hline
\end{tabular}
\end{center}
\caption{Time contribution of subroutines inside FSBM}.
\label{table:fsbm}

\end{table}
\end{comment}
Inspection of the call site shows that \verb|coal_bott_new| is called for each grid point but is nested within multiple conditionals and resides in the same loop as other processes (Listing \ref{lst:main}). Refactoring of this loop and other optimizations will be described in the section \ref{sec:implementation}.

\begin{lstlisting}[language=Fortran,label=lst:main, caption={Main loops over the grid points i,k,j calling multiple subroutines (simplified).}]
do j = jts,jte
  do k = kts,kte
    do i = its,ite
      if (T_OLD(i,k,j) > 193.15) then
        ! Nucleation
        call jernucl01_ks(...)
        ! Condensation
        if (...)
          call onecond1( ... )
        else if (...)
          call onecond2( ... )
        endif

        ! Collisions
        if (TT > 223.15) then
          call coal_bott_new( ... )
        endif
      endif
    enddo
  enddo
enddo
\end{lstlisting}

\section{Experimental Setup}\label{sec:setup}

The work in this paper is computed on the Perlmutter supercomputer at NERSC. Each node contains a single 2.45 GHz AMD EPYC 7763 (Milan) CPU with 64 cores and 4 NVIDIA A100 (Ampere) GPUs. Each GPU contains 40 or 80GB of HBM with a bandwidth of 1555 or 1935 GB/s, 108 streaming multiprocessors (SM), and achieves a peak performance of 9.7 TFLOP/s in double-precision, 19.5 TFLOP/s in single-precision.

As briefly mentioned, our test case is the standard CONUS-12km dataset which simulates thunderstorms on the Continental United States (CONUS) on a 425 \texttimes\ 300 \texttimes\ 50 grid with 12 km horizontal grid spacing. We run this case with a time step of 5 seconds for 10 simulation minutes. Throughout Section \ref{sec:approach}, all simulations are run with 16 MPI tasks, 1 OpenMP thread per task, and 1 GPU per task for the offloaded version. Section \ref{sec:multiple} describes cases with multiple GPUs per task.

Typically, WRF is compiled using GNU compilers on Perlmutter (option 35 in WRF's \verb|configure| support routine), loaded by the PrgEnv-gnu module. However, performance of the compiler versions available on the system is not optimal with GPU OpenMP offload and the latest version there is incompatible with the HPE-provided netCDF and HDF5 packages. Thus in this work we use the NVIDIA NVHPC compilers, loaded using the module PrgEnv-nvidia, which provides \verb|nvfortran| and \verb|nvc|. The corresponding \verb|configure| option in WRF is option 4 (pgf90/gcc). The complete configuration is shown in Table \ref{table:compilers}. Note that we use the HPE/Cray compiler wrappers \verb|ftn| and \verb|cc| which call the NVIDIA compilers once they are loaded.

\begin{table}[h!]
\begin{center}
\begin{tabular}{ c c   }
\hline
 Compilers & NVHPC 23.9 \\
 \hline
 Compiler flags & -pg -mp=gpu -target-accel=nvidia80 \\
 & -lvhpcwrapnvtx \\
 \hline
 Environment variables & NV\_ACC\_CUDA\_STACKSIZE=63336 \\ 
 & NV\_ACC\_CUDA\_HEAPSIZE=64MB \\
 \hline
\end{tabular}
\end{center}
\caption{Configuration of WRF on Perlmutter}.
\label{table:compilers}

\end{table}

\section{Approach}\label{sec:approach}
\subsection{Codee}
%Manuel
%Manuel[Helen added a paragraph about Codee below from the Codee training event announcement. Noticed there is already a short intro in Section I Introduction, so can do: Briefly in Introduction, and more details in here.  Manuel can decide]
%Manuel

Codee\cite{Codee} is a programming development tool for Fortran/C/C++ that facilitates the development of modern, parallel codes for multicore CPUs and GPUs using OpenMP and OpenACC. Leveraging the first Open Catalog of Best Practices for Modernization and Optimization\cite{OpenCatalog}, Codee identifies opportunities for improvement and provides detailed guidance on how to effectively exploit them. A standout feature is its ability to insert OpenMP and OpenACC directives, enabling even novice programmers to write parallel code for CPUs and GPUs in Fortran/C/C++. Additionally, Codee helps developers uncover hidden bugs, avoid introducing new ones, and pinpoint code optimization suggestions. As a result, Codee facilitates the maintenance and optimization of large Fortran/C/C++ codes, while ensuring code correctness and reliability. Codee also has the ability to automatically rewrite Fortran code to enforce Fortran modernization best practices, which is strongly recommended by experts before starting code optimization efforts.

%Manuel
%ManuelThis is how we set up Codee on Perlmutter...
%Manuel
On Perlmutter, Codee can be loaded with \verb|module load codee|.
To start, we first capture the compilation flags of all the WRF files with the \verb|bear| tool, which intercepts the actual compiler invocations while building WRF. Next, the source files are analyzed by invoking the Codee command \verb|screening| with this JSON file as an input, as shown in Listing \ref{lst:bear}.

\begin{lstlisting}[language=bash, caption=Commands for setting up Codee with WRF, label=lst:bear]
// Capture compilation flags in JSON file
bear -- ./compile -j 8 wrf

// Codee Screening report of WRF
codee screening --config compile_commands.json

// Codee Checks report of WRF
codee checks --config compile_commands.json

// Apply Codee AutoFix using OpenMP offload
codee rewrite --offload omp --in-place \
module_mp_fast_sbm.f90:6293:4 \
--config compile_commands.json
\end{lstlisting}
Due to the size of the WRF codebase, this process will take several hours. Once finished, we can use Codee \verb|checks| to list the checkers of the Open Catalog that apply to WRF, or to a specific file/subroutine. Finally, the examples above show how to instruct Codee to \verb|rewrite| a loop of FSBM in-place by annotating it with OpenMP offload directives.

%Manuel
%Manueln Listing \ref{lst:fix} we instruct Codee to find potential loops that can be offloaded. After verifying the output, we then apply the change in-place.
%Manuel
%Manuel
%Manuel\begin{lstlisting}[language=bash, caption=Commands for setting up Codee with WRF, label=lst:fix]
%Manuel// Codee Checks report of WRF
%Manuelcodee checks --config compile_commands.json
%Manuel
%Manuel// Apply Codee AutoFix using OpenMP offload
%Manuelcodee rewrite --offload omp --in-place \
%Manuelmodule_mp_fast_sbm.f90:6293:4 \
%Manuel--config compile_commands.json
%Manuel\end{lstlisting}
%Manuel

\subsection{Offloading with OpenMP}
Since version 4.0, OpenMP contains a set of directives for GPU kernel generation for both C/C++ and Fortran. These directives can manage parallelism and data transfers.  A commonly used combined construct to parallelize an n-nested loop is \verb|!$omp target teams distribute parallel do collapse(n)|. 
In the context of NVIDIA GPUs, the "target teams distribute" clause distributes CUDA thread blocks (each with 128 threads by default) over the loop iterations, and the "parallel do" clause assigns each thread to an iteration. The "collapse" clause combines nested loops into a single one to enable more work to be assigned to threads. By default, when entering an offloaded region, arrays are transferred to the device but scalar variables become \verb|firstprivate| for each thread. 

To manage host-device data transfers, OpenMP provides explicit mapping clauses; for example, \verb|map(to: A)| and \verb|map(from: A)| copies \verb|A| from the host to device and device to host, respectively. These constructs are essential in ensuring the least amount of data transfers, since by default OpenMP always performs data transfers when entering or exiting an offloading region regardless of necessity.

\section{Implementation}\label{sec:implementation}

\subsection{Lookup optimization and Codee}
Further inspection of the subroutine \verb|coal_bott_new| reveals that a significant amount of time is spent an inner subroutine \verb|kernals_ks|. 
The basic structure of this subroutine is shown in Listing \ref{lst:kernals}.

\begin{lstlisting}[language=Fortran, label=lst:kernals,caption={A typical loop nesting inside the collision kernal routine.}]
do j = 1,nkr
 do i = 1,nkr
  ckern_1 = ywls_750mb(i,j,1)
  ckern_2 = ywls_500mb(i,j,1)
  cwls(i,j) = (ckern_2+(ckern_1 - ...

  ckern_1 = ywlg_750mb(i,j,1)
  ckern_2 = ywlg_500mb(i,j,1)
  cwlg(i,j) = (ckern_2+(ckern_1 - ...
  
  ! 18 more arrays...
 enddo
enddo
\end{lstlisting}
These loops iterate over the \verb|nkr| $\times$ \verb|nkr| bins (in the current version, \verb|nkr| = 33), with each array on the left representing the interaction between two particle types; for example, the array \verb|cwls| refers to water (l) and snow (s). 
Thus, for each MPI task, the total amount of work for calls to \verb|coal_bott_new| is $O(mnkb^2)$, where $m,n,k$ are the number of grid points in each spatial direction, and $b$ the number of mass bins. 
Once all 20 of these collision arrays are filled, they are read later from other subroutines called within \verb|coal_bott_new|. 

These collision arrays were originally declared as global variables which prevents a simple parallelization of the 3 grid-level loops in Listing \ref{lst:main}, since they would be modified by different threads. However, applying Codee offloading directives reveals that there are actually no logical dependencies between grid points or array elements (Listing \ref{lst:codee}). Specifically, Codee applies a vectorization clause to the inner loop, and an OpenMP parallel and data offload clauses for the outer loop. The loop clauses imply that there are no loop-carried dependencies between the different iterations, and the \verb|map(from: ...)| clause implies that \verb|kernals_ks| in fact overwrites the collision arrays each time it is called and makes no use of previous values. Note that here we are not actually  using these directives themselves (since it would be too fine-grained); we are using the dependency analysis capability of Codee to gain insight on the loop structure.
\begin{lstlisting}[language=Fortran, label=lst:codee,caption={Directives for kernals\_ks applied by Codee.}]
! Codee: Loop modified
!$omp target teams distribute &
!$omp parallel do &
!$omp private(n) &
!$omp map(from: cwlg, cwls, ...
do j = 1,nkr
 ! Codee : Loop modified
 !$omp simd
 do i = 1,nkr
  ckern_1 = ywls_750mb(i,j,1)
  ckern_2 = ywls_500mb(i,j,1)
  cwls(i,j) = (ckern_2+(ckern_1 - ...

  ckern_1 = ywlg_750mb(i,j,1)
  ckern_2 = ywlg_500mb(i,j,1)
  cwlg(i,j) = (ckern_2+(ckern_1 - ...
  
  ! 18 more arrays...
 enddo
enddo
\end{lstlisting}

Based on this information, we completely removed the \verb|kernals_ks| subroutine and the global collision arrays \verb|cw**|, and instead compute each individual entry as needed when requested in other subroutines. This was done by writing new functions for each collision array which accepts the two indices as arguments, as shown in Listing \ref{lst:pure}.
Any subsequent access to, say \verb|cwlg(i,j)| is replaced by the function call \verb|get_cwlg(i,j,...)|. With this modification, there are no longer any shared states between different grid points in \verb|coal_bott_new| and parallelization is now straightforward. We also observe significant performance improvement from this change alone due to two main reasons: 1) not all 20 collision arrays are used, and 2) not every entry of an array is used. The speedups for \verb|fast_sbm| itself as well as the whole program are shown in Table \ref{table:noks}. Here and in the following tables, "current speedup" refers to the speedup compared to the previous version while "cumulative speedup" compares the current version to the version where the subroutine was first measured. These were calculated based on the time spent per time step of WRF.

\begin{lstlisting}[language=Fortran, label=lst:pure,caption={Example functions for computing an individual entry of each collision process.}]
pure real function get_cwlg(i,j, ...)
pure real function get_cwls(i,j, ...)
\end{lstlisting}

\begin{table}[h!]
\begin{center}
\begin{tabular}{  c c c  }
\hline
 &   Current speedup & Cumulative speedup  \\ 
 \hline
  %\verb|coal_bott_new| loop  &  & \\
  \verb|fast_sbm| & 1.83x  & 1.83x \\
  Overall & 1.42x & 1.42x \\
 \hline
\end{tabular}
\end{center}
\caption{Speedups of the FSBM routine and the whole program due to removal of kernals\_ks}.
\label{table:noks}
\end{table}

\subsection{OpenMP offloading}
As seen in Listing \ref{lst:main}, \verb|coal_bott_new| resides within large grid-level nested loops which also contain calls to other complex subroutines. To aid in programming efforts, we perform a loop fission to isolate \verb|coal_bott_new| by saving the states of variables before it was originally called in the main loop. We then finally apply the OpenMP offload directive on the outer loops (Listing \ref{lst:offload}). Here the predicate array \verb|call_coal_bott_new| stores the branching information from the original loops.

\begin{lstlisting}[language=Fortran, label=lst:offload, caption={Loops calling the collision subroutine isolated from the loops in Listing \ref{lst:main}.}]
!$omp target teams distribute &
!$omp parallel do collapse(2)
do j = jts,jte
  do k = kts,kte
    do i = its,ite
      if (call_coal_bott_new(i,k,j)) then
          call coal_bott_new( ... )
      endif
    enddo
  enddo
enddo
\end{lstlisting}
Note that, at this stage, we had to limit the collapse to 2 levels after encountering a runtime CUDA memory error due to stack overflow, which we later found to be caused by the large number of automatic arrays inside \verb|coal_bott_new|. The speedups from this offloading is shown in Table \ref{table:coal}. Here we also add a measurement for the isolated \verb|coal_bott_new| loop.

\begin{table}[h!]
\begin{center}
\begin{tabular}{  c c c  }
\hline
 &   Current speedup & Cumulative speedup  \\ 
 \hline
  \verb|coal_bott_new| loop  & 6.47x & 6.47x \\
  \verb|fast_sbm| & 1.54x  & 2.67x \\
  Overall & 1.33x & 2.09x \\
 \hline
\end{tabular}
\end{center}
\caption{Speedups of the collision loop, the FSBM routine and the whole program from offloading the outer 2 grid-level loops}.
\label{table:coal}
\end{table}

\subsection{Further optimization}
To avoid the aforementioned error on the GPU, we first increased the stack limit by setting the environmental variable \verb|NV_ACC_CUDA_STACKSIZE| to 65536 (measured in bytes). Next, we avoid the use of automatic arrays inside \verb|coal_bott_new| by creating allocatable arrays in a separate module, and then using pointers to refer to their slices. For comparison, Listing \ref{lst:coal1} shows the original declaration, and Listing \ref{lst:coal2} shows the modified version.
In the modified version, these arrays now point to slices of the external arrays corresponding to the grid point that is calling the subroutine.

\begin{lstlisting}[language=Fortran, label=lst:coal1, caption={Original declaration of the collision routine}]
subroutine coal_bott_new (...)
implicit none
!$omp declare target
! arguments...

! local variables
real :: fl1(33), fl2(33), fl3(33), ...
real :: g1(33), g2(33,icemax), g3(33), ...
    
\end{lstlisting}

\begin{lstlisting}[language=Fortran, label=lst:coal2,caption={Modified declaration which uses pointers to external arrays which are indexed by the grid point Iin, Kin, and Jin at which the subroutine is called.}]
subroutine coal_bott_new(Iin,Kin,Jin, ...)
use temp_arrays
implicit none
!$omp declare target
! arguments...

! local variables
real, pointer :: fl1(:),fl2(:),fl3(:), ...
real, pointer :: g1(:),g2(:,:),g3(:), ...

fl1 => fl1_temp(:,Iin,Kin,Jin)
fl2 => fl2_temp(:,Iin,Kin,Jin)
fl3 => fl3_temp(:,Iin,Kin,Jin)
g1 => g1_temp(:,Iin,Kin,Jin)
g2 => g2_temp(:,:,Iin,Kin,Jin)
g3 => g3_temp(:,Iin,Kin,Jin)

\end{lstlisting}
Here, the \verb|*_temp| arrays are declared in a separate module \verb|temp_arrays| which contains a subroutine to allocate  them once at the start of the simulation using the appropriate OpenMP data directives. For instance, \verb|fl1_temp| is allocated on the GPU through \verb|!$omp declare target (fl1_temp)| and \verb|!$omp target enter data map(alloc: fl1_temp)|.
While this uses more space overall (these arrays have to be allocated for all grid points at once and not only for currently active threads), it allows a full \verb|collapse(3)| of the main loops. The resulting speedups are shown in Table \ref{table:ptr}.

\begin{figure}
    \centering
    \includegraphics[width=0.99\linewidth]{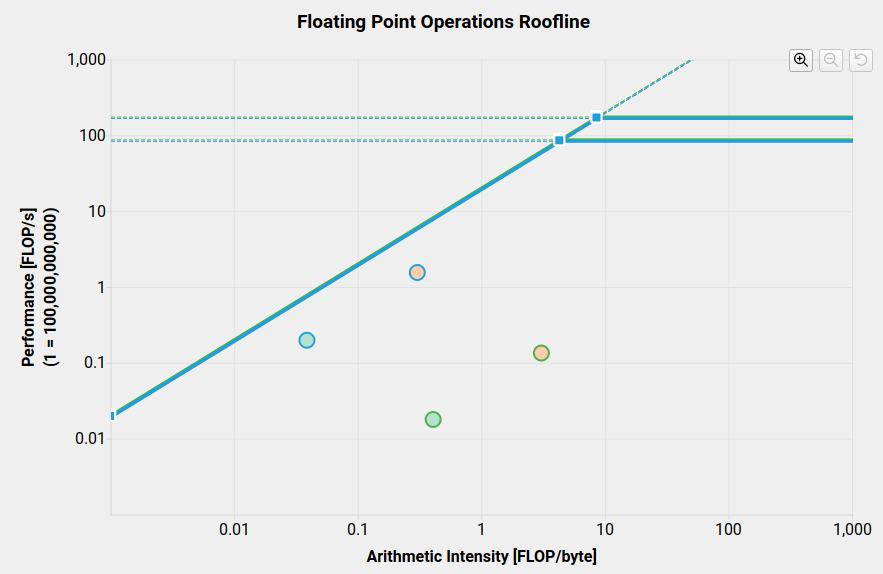}
    \caption{The solid lines form rooflines, with the top horizontal line for single precision and the bottom one for double precision. The green and brown circles at the bottom are the observed values with single and double precisions, respectively, when collapsing the two outermost loops. The pair of points above are when collapsing three loops.}
    \label{fig:roofline}
\end{figure}

\begin{table}[h!]
\begin{center}
\begin{tabular}{  c c c  }
\hline
 &   Current speedup & Cumulative speedup  \\ 
 \hline
  \verb|coal_bott_new| loop  & 10.3x & 66.6x \\
  \verb|fast_sbm| & 1.12x  & 2.99x \\
  Overall & 1.05x & 2.20x \\
 \hline
\end{tabular}
\end{center}
\caption{Speedups resulting from a full collapse of the grid-level loops through the removal of automatic arrays}.
\label{table:ptr}
\end{table}
To gain more insight into the performance characteristics, we used the NVIDIA Nsight Compute profiling tool, \verb|ncu|.
\begin{comment}
On Perlmutter, it is recommended to run \verb|ncu| for a single MPI task and disable dcgmi profiling. This is done through the bash script in Listing \ref{lst:ncu}. 

\begin{lstlisting}[language=bash, label=lst:ncu,caption={Wrapper script for Nsight Compute for MPI rank 3.}]
#!/bin/bash
if [[ ${SLURM_PROCID} == "3" ]] ; then
  dcgmi profile --pause
  ncu --set full  --target-processes all \
  --kernel-id :::2 -f $@
  dcgmi profile --resume
else
  $@
fi

\end{lstlisting}
\end{comment}
The generated roofline for the GPU versions (collapse twice and three times) is shown in Figure \ref{fig:roofline}. This plot reveals that using a full collapse significantly improves the performance of the loop, pushing it closer to the memory roofline. At the same time, using more threads decreases the arithmetic intensity, likely due to increased memory traffic as a result of a higher occupancy (see below) and more register spilling from having fewer registers per thread. These key details from Nsight Compute are shown in Table \ref{table:ncu}. Here, we see a significant reduction in kernel runtime and a sharp increase in occupancy from 4.63\% to 35.67\%, but lower cache hit rates and more transactions to global memory.

\begin{table}[h!]
\begin{center}
\begin{tabular}{ c c c  }
\hline
 Metric  & collapse(2) & collapse(3) w/ pointers  \\ 
 \hline
 Time (ms) & 335.85 & 29.11 \\
 Achieved occupancy (\%) & 4.63 & 35.67 \\
 L1/TEX hit rate (\%) & 84.82 & 61.43 \\
 L2 hit rate (\%) & 95.84 & 69.28 \\
 Writes to DRAM (GB) & 0.785 & 4.290 \\
 Reads from DRAM (GB) & 0.654 & 10.24 \\
 \hline
\end{tabular}
\end{center}
\caption{Comparison of metrics from Nsight Compute for the two offloaded codes}.
\label{table:ncu}
\end{table}

\section{Further evaluation}\label{sec:multiple}
\subsection{Using multiple MPI ranks per GPU}\label{sec:mps}
\begin{figure}
    \centering
    \includegraphics[width=0.95\linewidth]{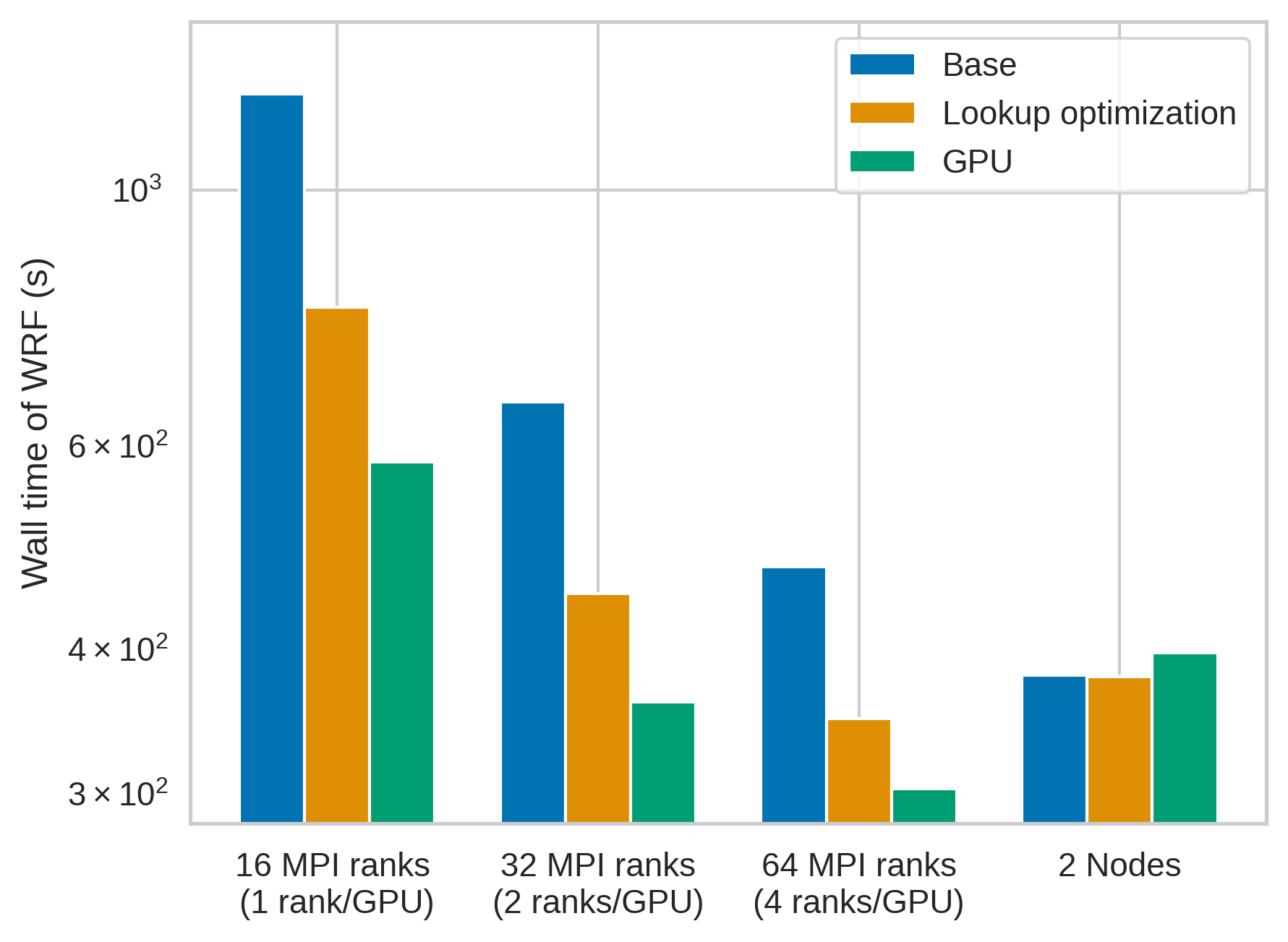}
    \caption{Total elapsed time for different versions of the code. For the GPU version, the number of GPUs is fixed to 16. In the rightmost group, the CPU codes run on 256 cores while the GPU code runs on 40 cores and 8 GPUs.}
    \label{fig:benchmark}
\end{figure}

In practice, we typically use more than 16 MPI ranks on most datasets. This section evaluates the performance of the program when we fix the number of GPUs to 16 on 4 nodes while increasing the number of CPU cores from 16 to 32 and 64. For each GPU, the (1/2/4) MPI tasks are distributed in a round-robin fashion. Wall clock times for 10-minute runs of different versions of the code shown in Figure \ref{fig:benchmark}. Time spent in I/O is also included in these measurements. Here, the GPU version refers to the final one with \verb|collapse(3)|.

For a more direct comparison, we also evaluate the case where the GPU and CPU codes each runs on 2 GPU/CPU nodes, respectively.  Here the CPU version runs on 256 MPI tasks, while the GPU version runs on 40 MPI tasks with 8 GPUs. Both versions still use 1 OpenMP thread per MPI task. The measurements are shown in the rightmost category in Figure \ref{fig:benchmark}. (We found that the current version of the code is limited to 5 MPI tasks per GPU, beyond which we encounter a CUDA memory error.) We summarize the timings and total speedups for the baseline and the final version of the code in Table \ref{table:final}.

\begin{table}[h!]
\begin{center}
\begin{tabular}{  c c c c }
\hline
 Configuration &   Time: & Time: & Total speedup  \\ 
 & baseline (s) & all optimizations (s) & \\
 \hline
  16 ranks & 1211.45 & 581.2 & 2.08x \\
  32 ranks & 655.1  & 360.1 & 1.82x \\
  64 ranks & 471.7 & 303.03 & 1.56x \\
  2 nodes & 379.8 & 397.1 & 0.956x \\
 \hline
\end{tabular}
\end{center}
\caption{Timing and speedup numbers for the baseline and final GPU version shown in Figure \ref{fig:benchmark}}.
\label{table:final}
\end{table}

\subsection{Output verification}
As a first pass in assessing the accuracy of the GPU versions, the tool \verb|diffwrf| (compiled as part of WRF) was used which reports bitwise differences between state variables in two input netCDF files. Sources of numerical differences include square root and fused multiply-add operations. When comparing the results of a 3-hour run (2160 time steps), 
we retain 3-6 digits for state variables such as velocities, temperature, and pressure, and 1-5 digits for microphysics variables. Note that while some quantities are double-precision, most in WRF are single-precision. To quickly evaluate the perturbation caused by each time step, we also used the \verb|-gpu=autocompare| flag which reports 6-7 digits of agreement.

\section{Discussion and Conclusion}\label{sec:conclusion}
This work examines performance improvements from optimizing parts of the WRF FSBM subroutine, serially as well as through OpenMP device offloading. When comparing the case with 16 MPI ranks and 1 GPU per rank on the CONUS-12km dataset, we observe a speedup of around 3x for the FSBM routine itself and a speedup of 2.2x for the whole program.  A limitation of the current code is memory: the GPU roofline plot (Figure \ref{fig:roofline}) reveals a low arithmetic intensity, especially for the fully collapsed version. The current implementation of FSBM involves loading many small arrays with length equal to $b$, the number of mass bins, and each grid point calls multiple subroutines that operate on these arrays. 
Thus, with our current parallelization strategy, accesses to these arrays are not coalesced but strided by $b$ elements. Additionally, due to the large number of these arrays and other scalar variables, the number of registers required per thread does limit occupancy. Manually limiting the register count resulted in significant speedup in the collapse(3) case, although further reduction beyond 64 appears to have no effect.

In the more realistic evaluations with multiple MPI tasks per GPU (Section \ref{sec:mps}), we still observe noticeable speedup when we increase the workload on each GPU by two to four times. 
One explanation is that FSBM calculations are not evenly distributed among the cores due to conditionals on the state variables which can vary spatially--many grid cells do not contain clouds, and thus require fewer calculations--so many GPUs were in fact underutilized in the 1 GPU/task case.  However, in the comparison using 2 CPU/GPU nodes, where total resources are made equal, the GPU version performs slightly worse. This is in part due to the high memory usage of the kernel which limits us to only 5 MPI ranks/GPU, and thus only 40 cores total. On the other hand, the CPU version with lookup optimization does not perform noticeably better than the baseline due to the dominating cost of MPI communication at 256 cores.
\begin{comment}
    
This is likely due to the loading FSBM-related datasets that occupy the same amount of space in \emph{each} domain which consequently scales linearly with the number of cores.
\end{comment}

Through the optimization process, we also demonstrated the combined use of runtime profilers and Codee to help accelerate code refactoring and identification of hot spots. Tools like gprof and Nsight Systems are valuable in prioritizing modules or subroutines to optimize, while the static analysis from Codee further delineates the structure/logic of specific loops in the code. In particular, the dependency analysis functionality of Codee enabled a quick restructuring of the collision arrays in \verb|kernals_ks| by confirming the lack of dependencies between grid points. This is especially helpful for programmers unfamiliar with the details behind the physics schemes used. 
Apart from the optimization in this particular example, we used the modernization checks from Codee to detect legacy constructs such as assumed-shape arrays and dummy argument intents in other subroutines like \verb|onecond|.

The loops calling condensation routines are currently being offloaded using a similar approach. Our next targets for offloading include other common microphysics routines like Thompson and P3, as well as scalar advection routines.

\section*{Acknowledgment}
This research used resources of the National
Energy Research Scientific Computing Center,
which is supported by the Office of Science of the
U.S. Department of Energy under Contract No.
DE-AC02-05CH11231.
Portions of the work were performed at Pacific Northwest National Laboratory (PNNL)—Battelle Memorial Institute operates PNNL under contract DEAC05-76RL01830.

\bibliographystyle{IEEEtran}
\bibliography{main.bib}

\vspace{12pt}

\end{document}